\documentclass[conference]{IEEEtran}
\usepackage{balance}
\usepackage{amsmath,amssymb,amsfonts}
\usepackage{algorithmic}
\usepackage{graphicx}
\usepackage{fancyhdr}
\usepackage[hyphens]{url}
\usepackage{float}
\usepackage{tikz}
\usepackage{listings}
\usepackage{subcaption,booktabs}
\usepackage{tabularx,booktabs}
\def\BibTeX{{\rm B\kern-.05em{\sc i\kern-.025em b}\kern-.08em
    T\kern-.1667em\lower.7ex\hbox{E}\kern-.125emX}}

\fancypagestyle{firstpage}{
  \fancyhf{}

  \fancyhead[C]{}
  \fancyfoot[C]{\thepage}
}

\title{A Flexible FPGA Accelerator for Convolutional Neural Networks}

\author{
  \IEEEauthorblockN{Kingshuk Majumder}
  \IEEEauthorblockA{
    Dept of CSA\\
    Indian Institute of Science\\
    Bengaluru, India\\
    kingshukm@iisc.ac.in
  }
  \and
  \IEEEauthorblockN{Shubham Nema}
  \IEEEauthorblockA{
    Dept of CSA\\
    Indian Institute of Science\\
    Bengaluru, India\\
    shubhamnema94@gmail.com
  }  
  \and
  \IEEEauthorblockN{Uday Bondhugula}
  \IEEEauthorblockA{
    Dept of CSA\\
    Indian Institute of Science\\
    Bengaluru, India\\
    udayb@iisc.ac.in
  }
}

\begin{document}
\maketitle
\thispagestyle{firstpage}
\pagestyle{plain}

\begin{abstract}
    Convolutional neural networks (CNNs) have become popular for machine perception
tasks on images and videos including image classification, object recognition,
and medical diagnosis.  The desire for high performance and power efficiency
while running CNNs has led to an explosion in specialized hardware accelerators.
FPGAs provide a promising alternative to ASICs and GPUs for flexible hardware
acceleration: reconfigurability makes them attractive on the Cloud for a wide
range of users whose requirements on precision and other aspects are likely to
be different.

Though CNNs are highly parallel workloads, in the absence of efficient on-chip
memory reuse techniques, an accelerator for them quickly becomes memory bound.
In this paper, we propose a CNN accelerator design for inference that is able to
exploit all forms of reuse available to minimize off-chip memory access while
increasing utilization of available resources.  The proposed design is composed
of cores, each of which contains a one-dimensional array of processing elements.
These cores can exploit different types of reuse available in CNN layers of
varying shapes without requiring any reconfiguration; in particular, our design
minimizes underutilization due to problem sizes that are not perfect multiples
of the underlying hardware array dimensions.

A major obstacle in the adoption of FPGAs as a platform for CNN inference is the
difficulty to program these devices using hardware description languages. Our
end goal is to also address this, and we develop preliminary software support
via a codesign in order to leverage the accelerator through TensorFlow, a
dominant high-level programming model. Our framework takes care of tiling and
scheduling of neural network layers and generates necessary low-level commands
to execute the CNN.

Experimental evaluation on a real system with a PCI-express based Xilinx VC709
board demonstrates the effectiveness of our approach. As a result of an
effective interconnection, the design maintains a high frequency when we scale
the number of PEs. The sustained performance overall is a good fraction of the
accelerator's theoretical peak performance, and to the best of our knowledge,
higher than previously published open designs with a similar setup.


\end{abstract}

\section{Introduction}
\label{sec:introduction}

Convolutional neural networks have enabled rapid progress in the fields of image
classification~\cite{resnet,resnext}, object
recognition~\cite{yolo,r-cnn,faster-r-cnn}, medical diagnosis from
scans~\cite{sayres19opthalmology}, and speech to text translation. All of these
fields fall into the field of machine perception involving images, video, and
speech. Besides delivering high accuracy, the simple and regular
characteristics of the computation have allowed system designers --- all the
way from architects, high-performance library and compiler developers, to
programming model designers --- to optimize and accelerate these computations in
turn leading to further innovation.  This has created a highly desirable
self-reinforcing feedback loop over the past seven years. CNNs have been
deployed on a vast range of platforms from data-centers to mobile phones.

While CNNs have high compute requirements, multi-core CPUs and GPUs have proved
to be efficient for training neural networks as a result of how well the
underlying matrix-matrix multiplication-like patterns have been optimized to run
at close to peak performance~\cite{goto2008toms,vanzee2015toms,lavin2015fast}.
However, most inference workloads have strict latency and power budgets.
Although ASICs can be used to implement high performance, power-efficient CNN
accelerators, but they are not very cost effective in the absence of a high
volume demand.

FPGAs provide a good compromise between cost effectiveness, performance and
power efficiency. With a significant amount of computing moving to the Cloud,
FPGAs are also attractive in that they could be customized to the varied
requirements of the multiple users a cloud server will have to support.
Such requirements could stem typically from precision, but also from other
aspects such as the CNN model itself and the problem sizes.  While reconfiguring
an FPGA while executing a particular user's workload may not be a practical
choice, providing a customized accelerator for a particular user is quite
appealing.

Designing an FPGA-based custom accelerator for CNNs is a difficult task, and
domain experts obviously should not have to think about hardware-specific
complexities involved. CNNs have abundant parallelism and the performance of a
hardware implementation, with sufficiently high compute power, could be
predominantly limited by available off-chip bandwidth if on-chip date reuse is
not effective. CNNs offer multiple data reuse opportunities such as input
feature map reuse along output channels, weight reuse within an input-output
channel pair, partial output sum reuse along the input channel dimension, and
convolutional reuse of both inputs and partial sums.  Exploiting available data
reuse is essential for a high performance accelerator design.  In addition, CNNs
offer other optimization opportunities such as reduced precision computing and
sparsity in input and weights.  In this work, we primarily focus on data reuse
while keeping the option of utilizing sparsity and reduced precision computing
open for future work. We describe the design and evaluation of an FPGA-based CNN
accelerator. We also build out the corresponding software support to utilize the
accelerator by leveraging it from high-level programming models like TensorFlow.

The ability to exploit available data reuse opportunities is heavily influenced
by the arrangement of processing elements (PEs), the design of the on-chip
interconnect~\cite{maeri} connecting the PEs, and the associated dataflow
techniques~\cite{eyeriss}. Different layers of a CNN can have very different
shapes and can thus offer better data reuse along different dimensions.  To
effectively exploit data reuse along multiple dimensions, previous work has
explored two-dimensional arrays of processing elements (PEs)~\cite{eyeriss, tpu,
bitfusion}. The dimensions of these two-dimensional arrays must be carefully
chosen to reduce under-utilization due to problem sizes that are not multiples
of the underlying processor array dimensions.  An accelerator may use flexible
interconnect design~\cite{maeri} or more complex mapping
techniques~\cite{eyeriss} to improve utilization of processing elements and data
reuse, but the added complexity contributes to increased area and power
consumption.


Our design constitutes multiple cores, where each core is a one-dimensional
array of processing elements (PEs). Each core can perform a convolution
operation of a single CNN layer or tiles of it if the former does not fit within
the resource constraints.
One of the characteristics of our design is that the under-utilization due to
the cleanup part of the problem size is minimized due to the flexibility of
mapping to the 1-d array. A tile of an arbitrary shape can be linearized to map
to our 1-d PE array. All of this is achieved while not compromising on data
reuse available along multiple dimensions.

In summary, our contributions are as follows.
\begin{itemize}
  \item We develop a CNN accelerator architecture that comprises multiple cores
    where each core is a one-dimensional array of processing elements. We show
    that a carefully designed one-dimensional design can obtain better
    utilization compared to 2-d designs. Also, the architecture can
    independently scale with available bandwidth and compute resources of an
    FPGA.
  \item We show that the proposed accelerator exploits data reuse along all
    dimensions.
  \item We exploit known data access patterns of CNNs to design a scalable and
    lightweight interconnect (in terms of resources) for transferring inputs to
    PEs.
  \item We develop a software framework to automate the process of running
    CNNs on FPGA.
\end{itemize}

The rest of this paper is organized as follows. Section~\ref{sec:background}
provides the necessary background on CNNs and the computational characteristics
to the extent they are relevant in designing parallel accelerators.
Section~\ref{sec:arch} describes our core architecture in detail along with an
analysis on how it exploits resources and properties of the computations
targeted in Section~\ref{sec:reuse}. Section~\ref{sec:software} describes the
software stack to make the accelerator usable with high-level programming
models. Section~\ref{sec:evaluation} presents our experimental evaluation
Section~\ref{sec:related-work} describes related work, and
conclusions are presented in Section~\ref{sec:conclusions}.

\section{Background: Convolutional Neural Networks}
\label{sec:background}
In this section, we provide the relevant background on CNNs and the convolution
operation.

A CNN is a feed-forward neural network containing
multiple layers.  The input and output of a layer are three-dimensional tensors
(ignoring batching).  Each layer computes a convolution operation, an
elementwise activation operation and an optional max-pooling.

\begin{figure}[htb]
    \centering
    \includegraphics[width=\linewidth]{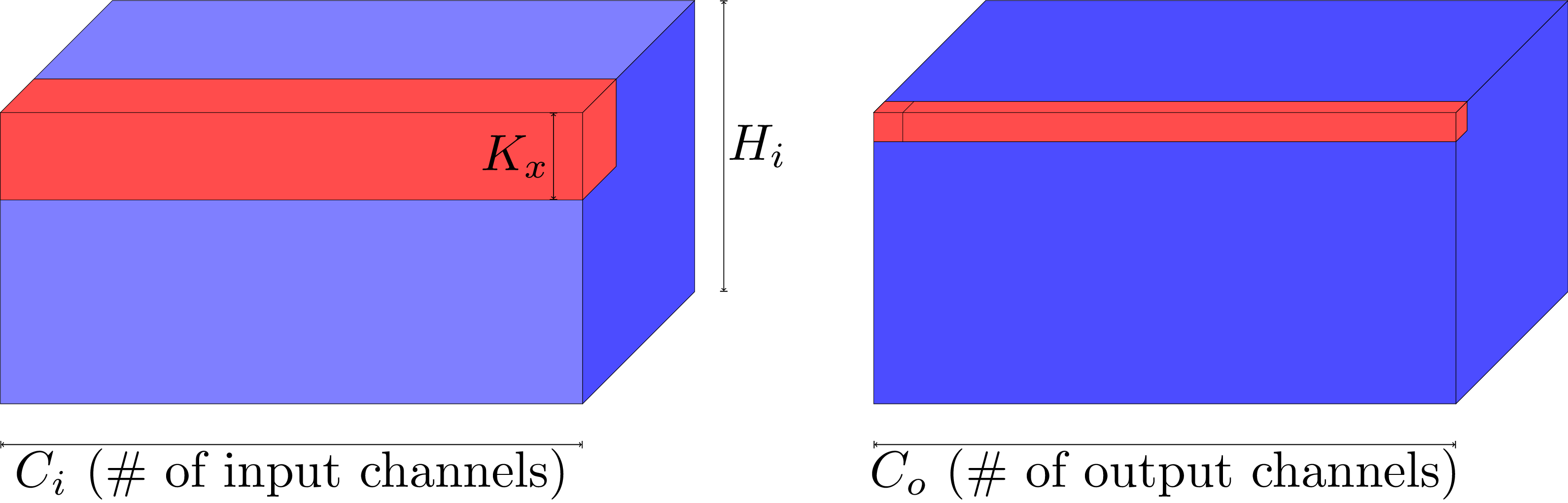}
    \caption{Convolution operation in a single layer of CNN.}
		\label{fig:conv}
\end{figure}

A convolution is the most compute heavy operation in a CNN.
Figure~\ref{fig:conv} shows a convolution layer with inputs and outputs. A
three-dimensional input tensor is convolved with a four-dimensional weight
tensor to calculate the output. The calculation of a single output value can be
represented as follows:
\begin{equation} \label{eq:conv} O(c_o, y, x) =
\sum\limits_{k_x,k_y,c_i} I(c_i, y+k_y, x+k_x) * W(c_o, c_i, k_y, k_x),
\nonumber \end{equation}

where $I$ is the input tensor (also called the input feature map or the input
channels) of shape $(C_i, H_i, W_i)$, $O$ is the output tensor (output feature
map or output channels) of shape $(C_o, H_o, W_o)$, and $W$ is the weight tensor
of shape $(C_o, C_i, K_y, K_x)$.  $C_i$ and $C_o$ denote the number of input and
output channels respectively, while $W_i$ and $H_i$ are the width and height of
a single channel of the input, (similarly $W_o$ and $H_o$ for output channels)
and $K_x$, $K_y$ represent the convolution window size.

A single convolution operation exhibits the following types of data reuse:
\begin{enumerate}
    \item{\textbf{input reuse:}} Each input value contributes to the calculation
    of $K_x*K_y*C_o$ outputs (assuming unit stride);
    \item{\textbf{weight reuse:}} Each weight value is reused for calculating
    $W_o*H_o$ output values (corresponding to a single
    input($c_i$)-output($c_o$) channel pair);
    \item{\textbf{partial sum reuse:}} Each output is calculated as a multiply
accumulate operation of $K_x*K_y*C_i$ input and weight values. Hence, the
    partial sum (psum) is reused $K_x*K_y*C_i$ times during this operation.
\end{enumerate}
Depending on the shapes, some CNN layers could have a higher input and psum
reuse, while others could have a higher weight reuse. Strides greater than one
reduce input reuse.  In order to achieve a high utilization of PEs in the
computation of each CNN layer, an architecture must be able to maximize reuse
along all dimensions.

\section{Architecture}
\label{sec:arch}

\definecolor{codegreen}{rgb}{0,0.6,0}
\definecolor{codegray}{rgb}{0.5,0.5,0.5}
\definecolor{codepurple}{rgb}{0.58,0,0.82}
\definecolor{backcolour}{rgb}{0.95,0.95,0.92}

\lstdefinestyle{style_pseudocode}{
    backgroundcolor=\color{backcolour},
    commentstyle=\color{codegreen},
    keywordstyle=\color{blue},
    numberstyle=\tiny\color{codegray},
    stringstyle=\color{codepurple},
    basicstyle=\scriptsize,
    breakatwhitespace=false,
    breaklines=true,
    captionpos=b,
    keepspaces=true,
    numbers=left,
    numbersep=5pt,
    showspaces=false,
    showstringspaces=false,
    showtabs=false,
    tabsize=3
}

\lstset{style=style_pseudocode}

In this section, we describe the core architecture of our proposed accelerator
design along with a detailed discussion of its characteristics, strengths, and
limitations.

\begin{figure}[htb]
  \centering
  \includegraphics[width=0.8\linewidth]{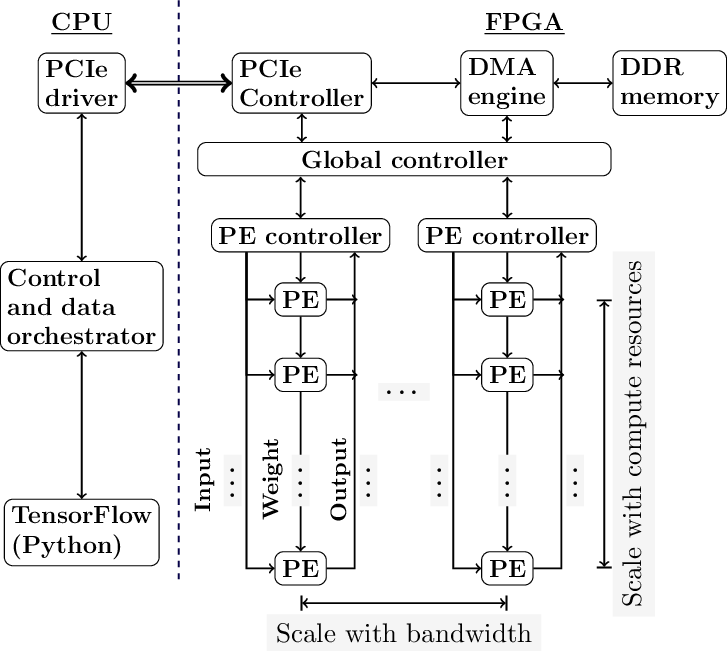}
  \caption{High-level design of the accelerator.}
  \label{tikz:fpga_pipeline}
\end{figure}

Figure~\ref{tikz:fpga_pipeline} gives a high level overview of our architecture.
Our accelerator is composed of one or more cores. Each core
contains multiple processing elements (PEs). A processing element is the
smallest compute resource in our design. Each processing element contains one
scalar fused-multiply-accumulate unit. In addition, PEs are provisioned with
local scratchpads to cache inputs, partial and final outputs. Each core contains
three interconnects to transfer inputs and weights to PEs and to read back
computed output. Each of these interconnects has the capacity to transfer one
value every cycle. A centralized controller within each core orchestrates
dataflow and generates control signals to schedule computation on the array of
PEs.

Based on the available bandwidth and compute resources, the exact number of
cores and PEs per core can be decided. As mentioned before, each interconnect
within a core can transferring one value every cycle.  Thus, maximum bandwidth
that a single core can utilize is limited by the capacity of these
interconnects.
Since, each core has a fixed peak bandwidth requirement, the total number of
cores is calculated by dividing available bandwidth (in the FPGA) with the
bandwidth requirement of one core. Number of PEs per core is limited by
available resources in the FPGA. Thus, our templated design scales with
bandwidth (by increasing number of cores) and compute (by increasing number of
PEs) independently.

\begin{lstlisting}[language=C++,caption=Pseudocode mimicking a single convolution
operation
scheduled in a core.,label={pseudocode},float]
// The two buffers below represent the local output buffer
// and partial output buffer of Wo*Ho PEs, each of which is
// of size Co.

float obuffer[Wo][Ho][Co];
float pobuffer[Wo][Ho][Co] = 0;

// The x and y loops are distributed among PEs, i.e., there
// are (Wo * Ho) active PEs.
for (x = 0; x < Wo; x++) {    // parallel
  for (y = 0; y < Ho; y++) {  // parallel
    float ibuffer[Kx][Ky];
    // Input channels are streamed sequentially.
    for (ci = 0; ci < Ci; ci++) {
      // Data is fetched into the input buffer.
      for (m = 0; m < Kx; m++) {
        for (n = 0; n < Ky; n++) {
          // Sx and Sy are stride along W and H axis.
          ibuffer[m][n] = ifmap[Sx * x + m][Sy * y + n][ci];
        }
      }
      // This loop nest is executed sequentially by one PE.
      for (i = 0; i < Kx; i++) {
        for (j = 0; j < Ky; j++) {
          for (co = 0; co < Co; co++) {
            // Final output.
            if (ci == Ci - 1 && i == Kx - 1
                             && j == Ky - 1) {
              obuffer[Wo][Ho][co] = pobuffer[Wo][Ho][co]
                + Filter[ci][i][j][co] * ibuffer[i][j];
            } else {
              // Partial output.
              pobuffer[Wo][Ho][co] += pobuffer[Wo][Ho][co]
                + Filter[ci][i][j][co] * ibuffer[i][j];
            }
          }
        }
      }
    }
    // Output is written back from output buffer.
    for (t = 0; t < Co; t++) {
      ofmap[x][y][co] = obuffer[co];
    }
  }
}
\end{lstlisting}

\subsection{Core}
Each core can compute convolution between a three dimensional input feature map
and a four dimensional weight tensor to calculate a three dimensional output.
Listing~\ref{pseudocode} shows the pseudocode for this computation. Loop nest
\textit{x, y} is parallelized by distributing it among PEs, each PE responsible
for executing one iteration of the loop nest. The operations within each PE can
be broken into three parts, input read (loops \textit{m} and \textit{n}),
computation (loops \textit{i, j, co, ci}) and output write back (loop
\textit{t}). The input read of the next channel and the output writeback of the
previous convolution are overlapped with the current execution to hide data
transfer latency. The convolution operation scheduled to a core must fit the
available resources. $W_o*H_o$ must be less than or equal to total number of PEs
in the core and the number of output channels $C_o$ must be less than the output
buffer size. The input is double buffered to hide data transfer latency; hence
it must be large enough to hold two $K_x*K_y$ windows of the input feature map.
A convolution operation that does not fit inside a core can be tiled along
$W_o$, $H_o$ and $C_o$ dimensions such that each tile fits the core. Such tiling
is done by the software runtime. The host side software communicates with the
core via PCIe to schedule convolution operations.

\subsection{Processing element}
\begin{figure}
  \centering
  \includegraphics[width=0.8\linewidth]{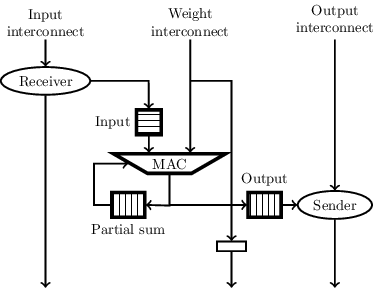}
  \caption{A Processing Element (PE).}
  \label{tikz:pe}
\end{figure}
Each core contains a one-dimensional array of processing elements.
Figure~\ref{tikz:pe} shows the internals of a PE.  A PE is the smallest compute
unit in our design. Each PE can perform one multiply-accumulate operation every
cycle. Partial outputs are cached within the PE and only the final outputs are
sent back. PEs do not have any controller and all control signals are sent by
core's controller. As discussed earlier, PEs overlap output writeback and input
read with current computation to avoid stalling. If the data transfer latencies
are less than the compute time, a PE will never stall.  Each PE contains one MAC
unit and scratchpad memory for inputs, partial sums and outputs. In addition, a
PE has a receiver and a sender node.  A receiver decides which data from the
input interconnect will be cached. A sender is responsible for reading the
output buffer and sending out the value via an interconnect.

Inputs and partial sums are cached to exploit temporal reuse within a PE.
Outputs are buffered to overlap the output read of a previous convolution with
current compute. As shown in Listing~\ref{pseudocode}, weights are reused across
PEs but there is no weight reuse within a PE. Hence, weights are not cached. A
PE calculates all $C_o$ feature channels corresponding to one pixel in the
output feature map using $K_x*K_y*C_i$ inputs cached in its local buffer. In our
current design, the input buffer is a 32-entry FIFO implemented using
distributed RAM.  Output and partial output buffers are 512-entry FIFOs realized
with block RAMs. As can be seen in Listing~\ref{pseudocode}, loop \textit{x} and
\textit{y} are distributed among PEs, and only $W_o*H_o$ PEs are active during a
convolution operation.  Hence, $W_o*H_o$ must be large for high utilization of
PEs within a core.  $W_o*H_o$ is also the total amount of weight reuse.

\subsection{Core controller}
The core controller is responsible for sending input and weights through the
interconnects, signaling PEs to perform the computation and reading back
outputs.  It receives commands from the software runtime with the convolution
parameters.  The controller then generates micro-instructions to control data
transfer and computation. It inserts appropriate stalls when the computation
time cannot completely hide data transfer time.

\subsection{Interconnect Design}
The performance of an accelerator heavily depends on the ability of the on-chip
interconnect to efficiently transport required data to the compute resource.
Without timely supply of data, these resources will stall, impacting overall
performance. Additionally, interconnects must be lightweight to ensure
sufficient FPGA resources are left for other parts of the design such as
compute.
In our design, each core has three interconnects to transport inputs, weights
and outputs respectively.
These interconnects can be classified as \emph{unicast}, \emph{multicast} and
\emph{broadcast}. The
output interconnect is responsible for collecting outputs from PEs and is a
unicast interconnect. In each cycle only a single PE is sending its output back
to the core controller.
The input interconnect is multicast as it needs to send one input to $K_x*K_y$
PEs. Each sent weight is required by all PEs, and hence, is transported via a
broadcast interconnect. All these interconnects are pipelined
to meet the desired frequency.
Next we discuss the mechanism by which each interconnect communicates with the
PEs.
The \emph{output interconnect} forwards read requests from the controller to all
the PEs. Each PE responds to this request by sending one value from its output
buffer back to the controller via the interconnect. Since the interconnect is
pipelined, PEs do not overwrite each others' output.

The \emph{weight interconnect} is the simplest of the three. Weights are sent by
the controller. Each PE uses the weight for updating its partial outputs and
in the subsequent clock cycle, forwards it to the next PE .

\begin{figure}[!htb]
  \centering
  \begin{subfigure}[b]{\linewidth}
  \centering
  \includegraphics[width=0.8\linewidth]{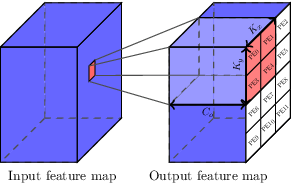}
  \caption{Receiver PEs (highlighted) for one input value. }
  \label{tikz:conv2}
  \end{subfigure}%
  \par\bigskip

  \begin{subfigure}[b]{\linewidth}
  \includegraphics[width=\linewidth]{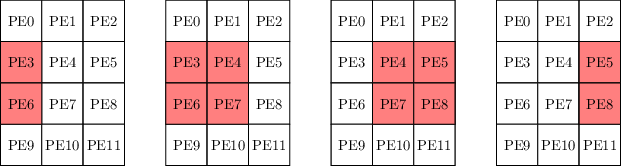}
  \caption{\textit{Receiver PEs} for consecutive input values of a row in the
    input feature map (unit stride, $2*2*C_o$ weights).}
  \label{tikz:SAPx}
  \end{subfigure}%
  \caption{Mapping to PEs.\label{fig:mapping}}
\end{figure}

\begin{table*}
  \begin{subtable}{0.4\linewidth}
  \centering
  \begin{tabularx}{\linewidth}{cc}
    \toprule
    \textbf{Command} & \textbf{Operation}                                 \\
    \midrule
    {13}                      & {Start (First input)}                     \\
    \midrule
    {0}                       & {No op}                                   \\
    \midrule
    {1}                       & {DilateX}                                 \\
    \midrule
    {2}                       & {ErodeX}                                  \\
    \midrule
    {3}                       & {ShiftX}                                  \\
    \midrule
    {4}                       & {RotateY}                                 \\
    \midrule
    {5}                       & {DilateY}                                 \\
    \midrule
    {6}                       & {ErodeY}                                  \\
    \midrule
    {7}                       & {ShiftY}                                  \\
    \midrule
    {8}                       & {Intermediate cmd for ShiftX}             \\
    \midrule
    {9}                       & {Intermediate cmd for DilateX}            \\
    \midrule
    {10}                      & {Intermediate cmd for ErodeX}             \\
    \midrule
    {11}                      & {Intermediate cmd for DilateY and ShiftY} \\
    \midrule
    {12}                      & {Intermediate cmd for DilateY and ShiftY} \\
    \bottomrule
  \end{tabularx}
  \caption{Commands sent to input interconnect from controller.}
  \label{tbl:start_cmd}
\end{subtable}
\hfill
\begin{subtable}{0.6\linewidth}
  \centering
  \begin{tabularx}{0.7\linewidth}{ccc}
    \toprule
    \textbf{Initial state} & \textbf{Command Change} & \textbf{Final state} \\
    \midrule
    {$RD=1$}               & {$3\mapsto8$}           & {$RD=0$}             \\
    \midrule
    {$RD=0$}               & {$8\mapsto3$}           & {$RD=1$}             \\
    \midrule
    {$RD=1$}               & {$1\mapsto9$}           & {$RD=1$}             \\
    \midrule
    {$RD=0$}               & {$9\mapsto1$}           & {$RD=1$}             \\
    \midrule
    {$RD=1$}               & {$2\mapsto10$}          & {$RD=0$}             \\
    \midrule
    {$RD=0$}               & {$10\mapsto2$}          & {$RD=0$}             \\
    \midrule
    {$-$}                  & {$4\mapsto4$}           & {$RD=LS$}            \\
    \midrule
    {$LS=1$}               & {$6\mapsto4$}           & {$LS=0$}             \\
    \midrule
    {$LS=1$}               & {$7\mapsto11$}          & {$LS=0$}             \\
    \midrule
    {$RD=1$}               & {$11\mapsto12$}         & {$RD=0$}             \\
    \midrule
    {$RD=0$}               & {$12\mapsto11$}         & {$RD=1, LS=1$}       \\
    \midrule
    {$LS=1$}               & {$5\mapsto11$}           & {$RD=1$}             \\
    \midrule
    {$-$}                  & {$13\mapsto14$}         & {$RD=1,LS=1$}        \\
    \midrule
    {$-$}                  & {$14\mapsto14$}         & {$RD=0,LS=0$}        \\
    \bottomrule
  \end{tabularx}
  \caption{State transition table for a receiver (inside a PE).}
  \label{tbl:cmd_trans}
\end{subtable}
\caption{Starting commands and command transition in the input interconnect
receivers.}
\label{tbl:interconnect_state}
\end{table*}

The \emph{input interconnect} is more complex because an input may be received
by a subset of PEs. Each input is cached by $K_x*K_y$ PEs for weight tensor of
size $K_x*K_y*C_i*C_o$.  The PEs that cache an input are not arbitrary and for
consecutive input values these receiving PEs change in a specific patterm.
Instead of using a general purpose multicast interconnect, we exploit this
pattern for a more lightweight design.  In order to understand how the PEs that
cache an input change with consecutive values along a row or column of input
feature map, we rearrange the one dimensional array of PEs as a logical two
dimensional array (as shown in Figure~\ref{tikz:SAPx}) of size $W_o*H_o$ for an
output of dimension $W_o*H_o*C_o$.

Each PE contains a receiver through which connected to the input interconnect. A
receiver contains two one bit flags, RD and LS, which together constitute is
state. RD indicates that the receiver is active and will read data from the
interconnect. LS indicates that it is the first PE in a row that contains active
receivers. In other words, LS indicates the start of rows that contain active
receivers.

We define the term \textit{receiver PEs}, for a given input value, as the set of
PEs that need to cache this value in their input buffer. For unit stride, each
input value (except at boundary) contributes to $K_x*K_y*C_o$ outputs as shown
in Figure~\ref{tikz:conv2}, where $C_o$ is the number of output channels.
Figure~\ref{tikz:SAPx} shows how the receiver PEs change for input values in a
row (assuming unit stride and $2*2*Co$ weights) of an input channel.  Receiver
PEs for the second input value \textit{dilates} one extra column to the right
compared to the first value.  From second to third value, the receiver PEs
\textit{shift} horizontally and from third to fourth value it \textit{erodes}
one column from left.  \emph{Note that, even with non unit strides, receiver PEs
  can never change more than one column in left or right direction for
consecutive input values of a row}.  Hence, change in receiver PEs for
consecutive input values of a row, can be expressed using three primitive
operations of \textit{dilation}, \textit{shifting} and \textit{erosion} along
x-axis.  Furthermore, for non unit strides, two adjacent inputs of a row may
have same receivers.

The input interconnect carries a 4-bit command in addition to data. The
receiver, upon receiving a command updates its state (RD and LS bits) and
forwards a new command to the next PE. Table \ref{tbl:cmd_trans} shows the
output command and updated state for a given input command and previous
state. Table \ref{tbl:start_cmd} lists all the commmands. Based on the
convolution window size and stride values, the controller sends
commands, such as \textit{DilateX} or \textit{ShiftY}, to update the set of
receivers that will cache the next input.
The resource overhead of input interconnect
consists of a 4 bit command bus to carry the command and the logic to implement
Table \ref{tbl:cmd_trans} in each PE, which is very small and scales linearly
with number of PEs.

\subsection{Support for different precisions}
\label{sec:precision} The current implementation of our architecture supports
both single precision floating point and 8-bit integer multiply with 32-bit
accumulate. In this section, we discuss certain unique challenges for the int8
configuration and how we address them.

In order to implement int8 operations (8-bit inputs, 8-bit weights and 32-bit
accumulation for output), we vectorize the PE. Thus, instead of doing one 32-bit
floating point multiply-accumulate operation, each PE now performs four 8 bit
integer multiplications and additions per clock. The software runtime packs a
set of four inputs into one 32-bit value. The dimensions we choose to pack into
a vector can have a significant impact on the design. One obvious choice is to
take sets of four batches and pack them together. This way, the core would
simultaneously be performing a convolution on four batch samples. The problem
with this approach is that int8 requires 32-bit accumulation. This means that
even though the inputs and weights are 8-bit wide, we will have to store the
partial outputs from all four batches in 32-bit precision. This increases our
block RAM usage by nearly four times.

\textbf{vPE:} To address this issue, we choose to pack four input channels into one vector
(and corresponding weight channels into one weight).
Each processing element calculates one dot product of the four packed inputs and
weights (while using 32 bit accumulators). This generates only one 32 bit
partial output (as opposed to four 32-bit partial outputs in the previous case).
The block RAM usage remains the same as with FP32, while we perform four
multiplies and four adds instead of one floating point multiply-accumulate. The
trade-off is that we exploit reduction parallelism along the input channel
dimension (four multiplies followed by an adder tree of three adds and an add to
accumulate) to generate the single reduced 32-bit output. Instead of calling
this 4-way parallel compute unit a PE, we refer to it as vPE here on.

\section{Data Reuse and Utilization}
\label{sec:reuse}
In this section, we describe how the proposed design exploits reuse in order to
achieve high performance and scale with resources.

Convolutional neural networks are highly parallel
workloads. Each output calculation is independent of others. In addition, there
are parallelization opportunities within the calculation of one output. Despite
this, performance cannot scale by merely increasing the amount of compute
resources. This is because the amount of performance that an accelerator can
achieve is limited by its ability to keep the compute units busy, i.e.
utilization of the available compute power. In order to achieve a high
utilization, an accelerator must exploit various data reuse opportunities
present in convolution operations.

On-chip data reuse is defined as the number of times a data participates in a
computation before being discarded. Section \ref{sec:background} described the
input, output and weight reuse available in a convolution operation. Each
convolution operation has an input reuse of $K_x*K_y*C_o$, weight reuse of
$W_o*H_o$ and output (partial sum) reuse of $K_x*K_y*C_i$.

In order to analyze different kinds of on-chip reuse and the utilization of
compute resources, we make following assumptions:
\begin{enumerate}
    \item The output feature map dimensions have been tiled such
      that $W_o \times H_o \leq $\textit{num PEs}, and $C_o \leq$ \textit{output
      buffer size} for an output tile of size $W_o*H_o*C_o$.  We also assume
      that $K_x*K_y \leq$ \textit{input buffer size}.
    \item The available bandwidth is enough to feed one input and weight to the
        core and read back one output from the core every cycle. We assume that
        bandwidth allocated for input cannot be traded for weight. This is
        because our design contains separate interconnects for inputs and
        weights that can send at most one input and one weight every cycle.
\end{enumerate}

We now describe how our core exploits reuse.  A unique property of our
accelerator is that it does not have any global scratchpad and all reuse is
either within a PE or due to sharing among PEs.  Based on this observation, we
classify the on-chip reuse into following two categories:
\begin{itemize}
  \item \emph{Intra-PE reuse} is the temporal reuse of data cached in
    local scratchpads of a PE. The amount of reuse is equal to the number of
    times a data is accessed from the local scratchpad before being overwritten.
  \item \emph{Inter-PE reuse} is the shared use of any data among multiple PEs.
    This constitutes any data that is sent to multiple PEs through the
    interconnects. The number of PEs that use a given data, defines the amount
    of its reuse.
\end{itemize}

\emph{Input reuse} is achieved through a mixture of inter-PE and
intra-PE reuse. Since each PE caches a $K_x*K_y$ window of inputs from an
input feature map, the number of PEs that cache the input is also equal to
$K_x*K_y$ (except boundary pixels which are required by less PEs). This is a
form of inter-PE reuse due to multiple PEs sharing same input.
In addition, once cached, each value in the $K_x*K_y$ window,
contributes towards updating $C_o$ partial sums, also cached in the PE.
Thus, each input value has an intra-PE reuse of $C_o$. In total, each input
value is reused $K_x*K_y*C_o$ times.
Since each weight sent through the interconnect is used by every active
PE, the \emph{weight reuse} is equal to the number of active PEs,
$W_o*H_o$. This reuse is completely inter-PE reuse. Since weights do
not have intra-PE reuse, they are not cached in PEs unlike inputs and
partial outputs.
Partial sums remain in a PE's local scratchpad until final outputs are
calculated. As can be seen in Listing~\ref{pseudocode}, each partial sum
participates in $K_x*K_y*C_i$ multiply accumulate operations before the
final outputs are ready. Thus, \emph{partial sum reuse} is completely
intra-PE reuse and is of factor $K_x*K_y*C_i$.

Thus, all of the available reuse for inputs, kernel and outputs, is either exploited
within local buffers of PEs or among multiple PEs without needing a global
buffer.
\subsection{Utilization of PEs}
\label{sec:utilization} The utilization of available compute resources places an
upper limit on the peak achievable performance. In this section, we develop an
analytical model to estimate the theoretical maximum utilization of PEs in a
single core of our design. We build this model with the previously defined
assumptions that the convolution operation is tiled to fit the core and that
there is enough bandwidth to supply one input and weight every cycle and
read back one output.

We categorize the utilization of PEs into two distinct types.
\begin{itemize}
    \item{\textbf{Spatial utilization:}} Since each PE calculates all channels
        corresponding to one output pixel, if the number of output pixels,
        $W_o*H_o$, is less than the number of PEs, then some PEs will
        remain inactive for the current operation.  We call the
        \textit{ratio of number of active PEs to the total number of PEs} as
        spatial utilization because it shows how many PEs are active for the
        duration of the convolution.

    \item{\textbf{Temporal utilization:}} We use double buffering of inputs and
        outputs to overlap compute time with data read/write time.  This ensures
        that PEs do not stall waiting for the input read or output writeback.
        However, if the input read or output writeback time is more than the
        compute time, the active processing elements will begin to stall
        waiting for the input or for output buffer to become available.  We call
        the ratio of \textit{compute time to the maximum of input read time and
        output writeback time} as the temporal utilization since it indicates
        the fraction of time the active PEs perform useful computation.  Since
        weights are used by all PEs and we previously assumed that we have a
        bandwidth to transfer at least one weight per cycle, PEs cannot
        stall due to weights.
\end{itemize}

Spatial utilization is defined by:
$$ U_s = \frac{W_o*H_o}{NUM\_PEs},$$ where $NUM\_PEs$ is the number of
processing elements in a core and the output feature map is of size
$W_o*H_o$.  Note that $U_s$ is always less than equal to one because we cannot
schedule a convolution operation on the core that has $W_o*H_o> NUM\_PEs$.

Listing~\ref{pseudocode} shows the number of computations sequentially executed
by one PE. Since we read one input per cycle, the time required to read one
channel of input feature map is $W_i*W_i$. Each PE caches $K_x*K_y$ window
of this feature map and updates the partial outputs. It takes $C_o*K_x*K_y$
cycles to update all the partial outputs using this window. Thus in order to
keep the PEs busy all the time, input read time must be less than the compute
time. In case input read time is more, the PEs will stall for the remaining
cycles waiting for inputs to arrive. Thus, $(C_o*K_x*K_y)/(W_i*W_i)$ gives
an upper bound on temporal utilization.
Similarly the time required to send back all outputs is $W_o*H_o*C_o$.
Hence, temporal utilization is given by the formula:
\[ {U}_t = \min\left(1, \frac{C_i*K_x*K_y*Co}{\max
\left(W_i*W_i*C_i,W_o*H_o*C_o\right)}\right).  \]

The upper limit on $U_t$ is reached when compute time is equal to or more
than the input/output read/write time.

The total utilization of a core is given by $U = U_s*U_t$ which is equal to
\begin{eqnarray}
U &= \min &\left( \frac{W_o*H_o}{num\_PEs}, \right. \nonumber  \\
&&
 \left.\frac{C_i*K_x*K_y*Co}{\max
\left(W_i*W_i*C_i,W_o*H_o*C_o\right)}\right). \nonumber
\end{eqnarray}

When the output feature map is larger than the total number of PEs, we tile it
 to fit the core.

\section{Software Framework}
\label{sec:software}
In this section, we describe the software stack that integrates the accelerator
into the TensorFlow toolchain.
\begin{figure}[!h]
    \centering
		\includegraphics[scale=0.8]{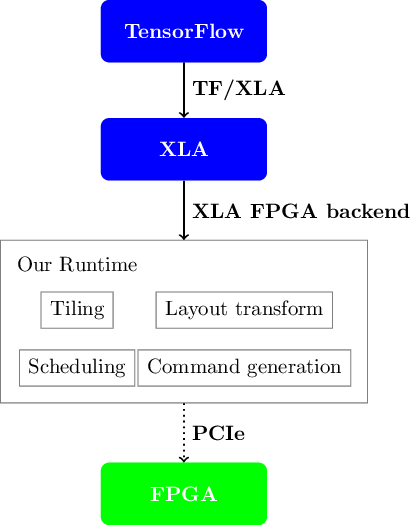}
    \caption{Software stack.}
    \label{tikz:software_stack}
\end{figure}

\begin{figure}[H]
    \centering
    \includegraphics[width=0.3\linewidth]{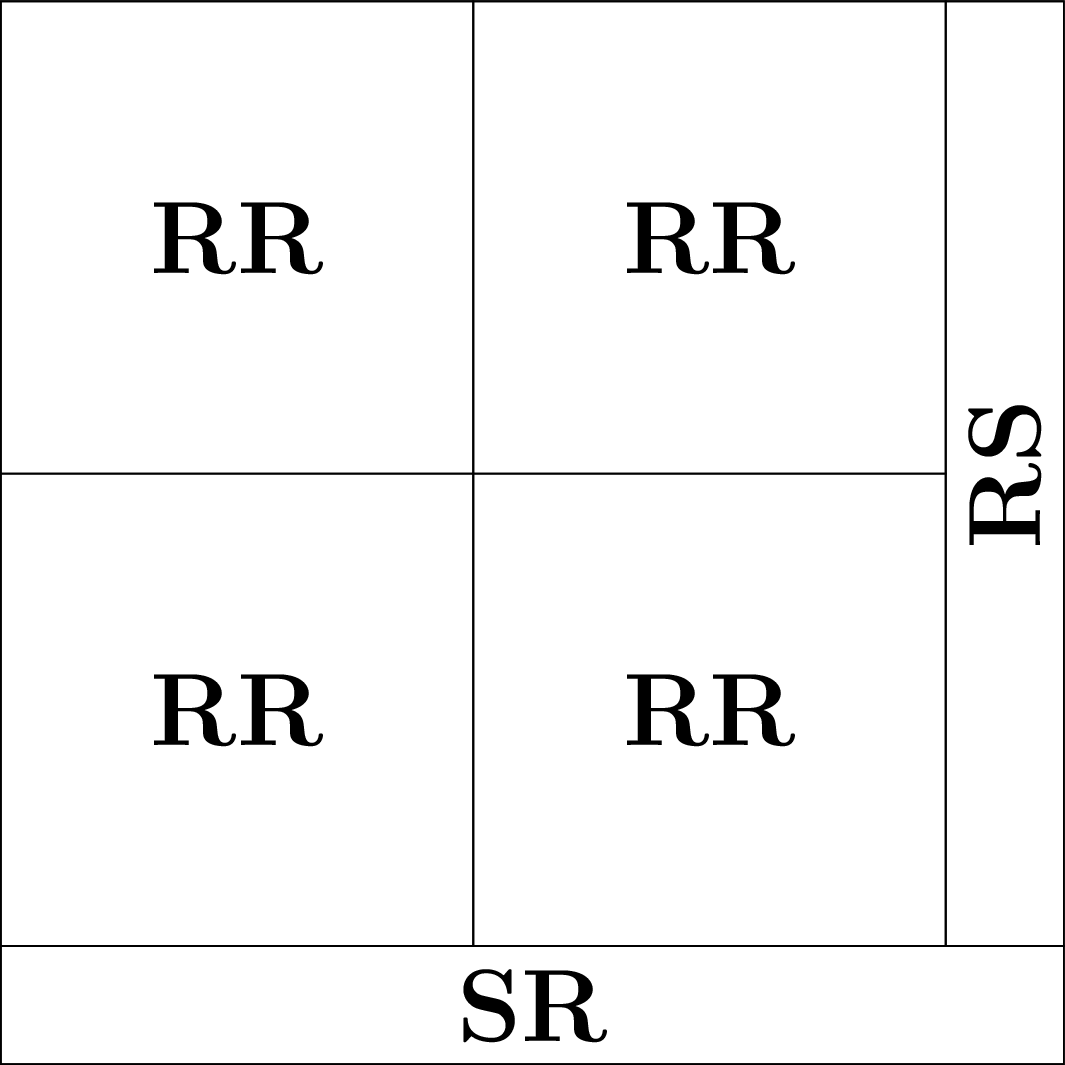}
		\caption{Tiling of output feature map in $H_o\times W_o$ plane.}
    \label{fig:tiling}
\end{figure}

\subsection{Mapping linearized shapes to reduce under-utilization}
Figure~\ref{fig:tiling} shows our tiling strategy. We tile the $W_o$ and
$H_o$ dimensions so that each tile is of size $T_{ox}*T_{oy}*C_o$. If
required, the output channel dimension, $C_o$, can also be tiled. However, in
our experiments, we did not have to tile the output channel dimension since the
output buffer was large enough to fit all output channels.  To make the tiling
implementation simpler, we always keep the number of PEs a perfect square. Our
architecture however has no restriction on the number of PEs.  Assuming that the
number of PEs is $P^2$, we try to fit as many $P\times P$ tiles as we can,
starting from the top left corner. In Figure~\ref{fig:tiling}, this is shown by
the tiles marked RR (R stands for regular-regular). This leaves us with two
partial tiles \textit{RS} (regular-small) and \textit{SR}.  Next, we attempt to
schedule the \textit{RS} tile.  If the number of output pixels in the
\textit{RS} tile is less than the total number of PEs in a core, we can schedule
the complete \textit{RS} tile.  Otherwise, we further divide it into smaller
sub-tiles. A sub-tile will have the same width, $W$ as the \textit{RS} tile.
Note that $W=(W_o \bmod P)<P$. Thus we are guaranteed that we will be able to
fit at least a sub-tile of size $W\times P$ in the core. We divide the
\textit{RS} tile along the $H_o$ dimension (height) into sub-tiles, where each
subtile is of size $(W_o \bmod P)*P$ except the last tile which may be of height
less than $P$.  Similarly, we break down the \textit{SR} partial tile into
subtiles along the $W_o$ dimension.  The sub-tiles that originate from
\textit{RS} and \textit{SR} can have a very skewed aspect ratio. For example, if
we have 64 PEs per core and $W_o=H_o=65$ then, the partial tile \textit{SR} will
be of size $1\times 64$, i.e., a one pixel wide column of output values. Our
one-dimensional PE array design does not put any constraint on the aspect ratio
of the scheduled tile. Hence, we can schedule the complete $1\times 64$ tile in
the core (which has 64 PEs).

\subsection{Software stack}
Figure~\ref{tikz:software_stack} provides an overview of our software stack. The
high-level description of a CNN is specified as a TensorFlow model, which is our
starting point.  Our runtime takes the input as an XLA HLO graph, an
intermediate representation on the path of TensorFlow compilation. The runtime
then performs the necessary data layout transformation on inputs, tiles each
convolution layer to fit on the FPGA, and generates low-level instructions to
schedule each tile on the FPGA.

Our design requires the layout of weights to be $C_i\times K_y\times K_x\times
C_o$, while TensorFlow uses a layout of $K_y\times K_x\times C_i\times C_o$.
Similarly, we need the input/output feature map layout to be $C\times W \times
H$, but TensorFlow's default layout is $W\times H \times C$ (ignoring the batch
dimension).  Our runtime helper performs the necessary layout transformation.
It then creates a schedule for executing tiles, packs the required input feature
channels and weights, along with instructions to execute the convolution
operation on the core, and sends this packet to the FPGA via PCIe. Upon
receiving the results, the runtime unpacks the output to the correct layout and
returns it to TensorFlow.

While the key contribution of this work is on the hardware side, our larger
longer term goal is to build HLS support to it through a dialect in
MLIR~\cite{mlir}, which is an intermediate representation to which TensorFlow is
moving, and which potentially other AI/ML compilers are likely to adopt.

\section{Experimental Evaluation}
\label{sec:evaluation}
In this section, we describe the experimental setup, the evaluation performed,
and an analysis of the results. All performance and execution time measurements
are from a real experimental system.

\subsection{Setup and Methodology}
We implemented our accelerator using Verilog. The experimental setup constitutes
a Xilinx VC709 FPGA evaluation board connected over PCI-ex 3.0 (via an x16
interface) on an Intel Xeon E5-2630 v3 server (Intel Haswell-based) with 64 GB
of DDR4-1600  RAM for experimental evaluation.  All reported numbers are with
Vivado 2019.1 being used for synthesis. Riffa~\cite{riffa} was used for host to
FPGA communication via PCIe. The maximum PCIe bandwidth achievable was 4GB/s in
each direction. For single precision floating point, we use Xilinx's
floating point IP. For int8 precision, we wrote custom multipliers in Verilog
using DSP blocks, and the adder tree (Section~\ref{sec:precision}) using LUTs.
The input buffer is a 32 entry FIFO and output and partial output buffers are
512 entry FIFOs. TensorFlow release version 1.12 was used for running the
models. All performance numbers are for inference with a \emph{batch size} of
$1$.

\begin{table}[tb]
  \centering
  \begin{tabularx}{0.95\linewidth}{l @{~} c @{~} c @{~} c @{~} c @{~} c @{~} c }
    \toprule
    {\bf Layer} & Output dims   & Inp ch. & Conv window & \multicolumn{3}{c}{Arithmetic}\\
                & ($H$ x $W$ x $C_o$) & ($C_i$) & $K_x$ x $K_y$ & \multicolumn{3}{c}{intensity} \\
		\cmidrule(lr){5-7}
& & & & Input & Weight & Output \\
    \midrule
    conv1\_1    & {224x224x64}  & 3   & {3x3} & 1152 & 100352 & 54        \\
    \midrule
    conv1\_2    & {224x224x64}  & 64  & {3x3} & 1152 & 100352 & 1152        \\
    \midrule
    conv2\_1    & {112x112x128} & 64  & {3x3} & 2304 & 25088  & 1152        \\
    \midrule
    conv2\_2    & {112x112x128} & 128 & {3x3} & 2304 & 25088  & 2304       \\
    \midrule
    conv3\_1    & {56x56x256}   & 128 & {3x3} & 4608 & 6272   & 2304       \\
    \midrule
    conv3\_2    & {56x56x256}   & 256 & {3x3} & 4608 & 6272   & 4608      \\
    \midrule
    conv3\_3    & {56x56x256}   & 256 & {3x3} & 4608 & 6272   & 4608     \\
    \midrule
    conv4\_1    & {28x28x512}   & 256 & {3x3} & 9216 & 1568   & 4608      \\
    \midrule
    conv4\_2    & {28x28x512}   & 512 & {3x3} & 9216 & 1568   & 9216     \\
    \midrule
    conv4\_3    & {28x28x512}   & 512 & {3x3} & 9216 & 1568   & 9216    \\
    \midrule
    conv5\_1    & {14x14x512}   & 512 & {3x3} & 9216 & 392    & 9216    \\
    \midrule
    conv5\_2    & {14x14x512}   & 512 & {3x3} & 9216 & 392    & 9216   \\
    \midrule
    conv5\_3    & {14x14x512}   & 512 & {3x3} & 9216 & 392    & 9216  \\
    \bottomrule
  \end{tabularx}
  \caption{VGG-16 convolution layer sizes\label{tab:vgg16}.}
\end{table}

\begin{table*}[tb]
\parbox{0.48\linewidth}{
  \centering
  \begin{tabularx}{\linewidth}{  l c  c  c  c }
    \toprule
    {\bf Resource}        & LUT & FF   & BRAM (18Kb) & DSP\\
    \midrule
    MAC unit              & 663 & 1073 & 0           & 2\\
    \midrule
    Input buffer          & 58  & 46   & 0           & 0\\
    \midrule
    Partial output buffer & 48  & 80   & 1           & 0\\
    \midrule
    Output buffer         & 48  & 80   & 1           & 0\\
    \midrule
    Misc                  & 73  & 110  & 0           & 0\\
    \midrule
    {\bf Total}           & 890 & 1389 & 2           & 2\\
    \bottomrule
  \end{tabularx}
  \caption{Resource usage of a single fp32 PE\label{tab:pe_usage_fp32}.}
}
\parbox{0.48\linewidth}{
  \centering
  \begin{tabularx}{\linewidth}{  l c  c  c  c }
    \toprule
    {\bf Resource}        & LUT & FF  & BRAM (18Kb) & DSP\\
    \midrule
    MAC unit              & 131 & 166 & 0           & 4\\
    \midrule
    Input buffer          & 79  & 85  & 0           & 0\\
    \midrule
    Partial output buffer & 48  & 80  & 1           & 0\\
    \midrule
    Output buffer         & 48  & 80  & 1           & 0\\
    \midrule
    Misc                  & 50  & 71  & 0           & 0\\
    \midrule
    {\bf Total}           & 356 & 482 & 2           & 4\\
    \bottomrule
  \end{tabularx}
  \caption{Resource usage of a single int8 vPE\label{tab:pe_usage_int8}.}
}
\end{table*}

\begin{table}[htb]
  \centering
  \begin{tabularx}{0.95\linewidth}{  l c  c  c  c }
    \toprule
    {\bf Resource}         & \multicolumn{4}{c}{\bf Available} \\
    \cmidrule(lr){2-5}
                           & LUT     & FF      & BRAM (36Kb) & DSP  \\
    \midrule
                           & 433,200 & 866,400 & 1,470       & 3,600\\
    \midrule
    {\bf Configuration} & \multicolumn{4}{c}{\bf Utilization} \\
    \cmidrule(lr){2-5}
                           & LUT     & FF      & BRAM (36Kb) & DSP  \\
    \midrule
    16 PEs                 & 6\%     & 5\%     & 5\%         & 1\% \\
    \midrule
    64 PEs                 & 16\%    & 13\%    & 8\%         & 4\%\\
    \midrule
    256 PEs                & 55\%    & 43\%    & 21\%        & 14\%\\
    \midrule
    324 PEs                & 69\%    & 54\%    & 26\%        & 18\%\\
    \bottomrule
  \end{tabularx}
  \caption{Hardware utilization for fp32 PEs\label{tab:util-fp32}.}
\end{table}

\begin{table}[tb]
  \centering
  \begin{tabularx}{0.95\linewidth}{  l c  c  c  c }
    \toprule
    {\bf Configuration} & \multicolumn{4}{c}{\bf Utilization}        \\
    \cmidrule(lr){2-5}
                            & LUT    & FF     & BRAM (36Kb) & DSP    \\
    \midrule
    256 vPEs                & 25.1\% & 16.7\% & 21.2\%      & 28.6\% \\
    \midrule
    324 vPEs                & 31\%   & 20.5\% & 25.9\%      & 36.1\% \\
    \midrule
    400 vPEs                & 38\%   & 25\%   & 31\%        & 45\%   \\
    \midrule
    625 vPEs                & 56.8\% & 37.5\% & 46.5\%      & 69.6\% \\
    \bottomrule
  \end{tabularx}
  \caption{Hardware utilization for int8 vPEs\label{tab:util-int8}.}
\end{table}

We ran our experiments for the convolution layers of
VGG-16~\cite{simonyan2014arxiv}.
Table~\ref{tab:vgg16} shows the sizes of all the convolution layers along with
their inherent arithmetic intensities (ratio of ops to the size of data).  We
used a batch size of one.  All results are with the design running at a 250~MHz
frequency.  We present here results with 32-bit floating point (fp32) and with
8-bit integer (int8) precision. While int8 is the commonly evaluated precision
for inference for accelerators, we evaluate fp32 as well here as the necessary
model (weights) were available and could be easily tested on the VGG model
available with TensorFlow. Getting valid weights and a model with 8-bit
inference would require more elaborate software support in conjunction with the
TensorFlow toolchain, and so for int8, we used synthetic weights. The measured
performance would be exactly the same as with real weights since we do not use
any data dependent optimization (such as exploiting sparsity to reduce
computation). In all cases, all our performance results are from measurement on
runs on a real system. They were verified for correctness against a reference
CPU implementation.

Although more complex state-of-the-art CNNs like ResNet~\cite{resnet},
ResNeXt~\cite{resnext}, and R-CNN~\cite{r-cnn} exist, we chose VGG so that the
experimentation could focus on the core primitive: the same performance
characteristics and insights carry over to convolution layers in other models
since we are really accelerating a ``kernel'' underlying convolutions as opposed
to something specific to VGG.

We performed all experiments for a single core, i.e., all the PEs were present
in a single core as opposed to being split across multiple cores (see
Section~\ref{sec:utilization}).

\begin{table*}[!htb]
  \begin{tabularx}{\linewidth}{cccc c c@{~~}c c@{~~}c c@{~~}c c@{~~}c}
    \toprule
    \multicolumn{4}{c}{Convolutional layer} & GOPs & \multicolumn{2}{c}{16 PEs @
    250 MHz} &
    \multicolumn{2}{c}{64 PEs @ 250 MHz} & \multicolumn{2}{c}{256 PEs @ 250 MHz}
    & \multicolumn{2}{c}{324 PEs @ 250 MHz}
    \\
    \cmidrule(lr){1-4} \cmidrule(lr){6-7} \cmidrule(lr){8-9}
    \cmidrule(lr){10-11} \cmidrule(lr){12-13}
    Height & Width & Input ch. & Output ch. & & Time (ms) & GFLOPS & Time (ms) &
    GFLOPS & Time (ms) & GFLOPS & Time (ms) & GFLOPS\\
    \midrule

    224 &224 & 3  & 64 &0.16 & 27.9 &5.79 & 26.7  &6.05    &26.3 &6.15   &26.4
        &6.13  \\
    224 &224 &64  & 64 &3.45 &491.8 &7.01 &123.3  &27.95   &31.3 &110.11 &26.4
        &130.44\\
    112 &112 &64  &128 &1.72 &238.8 &7.21 & 60.0  &28.71  &15.4 &111.71 &13.5
        &127.38\\
    112 &112 &128 &128 &3.45 &477.3 &7.22 &119.6  &28.80   &30.4 &113.39 &24.8
        &139.15\\
    56  &56  &128 &256 &1.72 &235.1 &7.33 & 59.2  &29.11   &16.2 &106.61 &13.7
        &126.13\\
    56  &56  &256 &256 &3.45 &470.0 &7.33 &117.9  &29.22   &31.7 &108.78 &26.8
        &128.46\\
    56  &56  &256 &256 &3.45 &470.0 &7.33 &117.9  &29.23   &31.7 &108.75 &26.9
        &128.30\\
    28  &28  &256 &512 &1.72 &233.4 &7.38 & 62.2  &27.70   &19.6 &87.71  &15.3
        &112.58\\
    28  &28  &512 &512 &3.45 &466.4 &7.39 &124.0  &27.78   &38.7 &89.09  &29.6
        &116.48\\
    28  &28  &512 &512 &3.45 &466.4 &7.39 &124.0  &27.78   &38.7 &89.09  &29.6
        &116.58\\
    14  &14  &512 &512 &0.86 &123.9 &6.95 & 38.3  &22.48   &10.6 &81.63  &10.6
        &81.24 \\
    14  &14  &512 &512 &0.86 &123.9 &6.95 & 38.3  &22.47   &10.5 &81.98  &10.5
        &81.81 \\
    14  &14  &512 &512 &0.86 &123.9 &6.95 & 38.3  &22.46   &10.5 &81.88  &10.5
        &81.84 \\
	\midrule
	\multicolumn{5}{l}{Theoretical peak performance (add/mul) (GFLOPS)} & & 8 & & 32 & &  128 & & 162 \\
	\multicolumn{5}{l}{Overall performance (GFLOPS)} & & 7.24 & & 27.23 & &  91.79
                                                   & & 108.1 \\
  \multicolumn{5}{l}{Max fraction of peak sustained} & & 92.3\% & & 91.3\% & &
    88.5\% & & 85.9\% \\

    \bottomrule
  \end{tabularx}
  \caption{Performance breakdown of each layer of VGG-16\label{tab:perf}. GFLOPS
  is fp32 GFLOPS. Batch size is 1}
\end{table*}

\begin{table*}[!htb]
  \begin{tabularx}{\linewidth}{cc@{~~}c@{~~}c @{~~} c @{~~} c@{~~}c c@{~~}c c@{~~}c c@{~~}c}
    \toprule
    \multicolumn{4}{c}{Convolutional layer} & GOPs &
    \multicolumn{2}{c}{256 vPEs @ 250 MHz} & \multicolumn{2}{c}{324 vPEs @ 250
    MHz} & \multicolumn{2}{c}{400 vPEs @ 250 MHz} & \multicolumn{2}{c}{625 vPEs @ 250 MHz} \\
    \cmidrule(lr){1-4} \cmidrule(lr){6-7} \cmidrule(lr){8-9}
    \cmidrule(lr){10-11} \cmidrule(lr){12-13}
    Height & Width & Input ch. & Output ch. & & Time (ms) & GOPS & Time (ms) &
    GOPS & Time (ms) & GOPS & Time (ms) & GOPS\\
    \midrule
    224 &224 & 3  & 64 &0.16 &13.46 & 11.99  & 13.56 & 11.91  & 13.51 & 11.95  & 13.40& 12.06   \\
    224 &224 &64  & 64 &3.45 &13.67 & 252.03 & 13.74 & 250.70 & 13.52 & 254.85 & 13.60& 253.31  \\
    112 &112 &64  &128 &1.72 &7.18  & 240.09 & 7.10  & 242.76 & 6.97  & 247.29 & 6.97 & 247.19  \\
    112 &112 &128 &128 &3.45 &8.00  & 430.66 & 7.04  & 489.32 & 7.15  & 481.66 & 7.28 & 473.26  \\
    56  &56  &128 &256 &1.72 &4.61  & 373.92 & 3.82  & 450.49 & 3.93  & 438.78 & 3.83 & 449.43  \\
    56  &56  &256 &256 &3.45 &8.23  & 418.58 & 7.04  & 489.25 & 5.95  & 578.66 & 4.18 & 823.84  \\
    56  &56  &256 &256 &3.45 &8.24  & 418.02 & 7.06  & 488.14 & 5.93  & 581.39 & 4.20 & 820.90  \\
    28  &28  &256 &512 &1.72 &5.09  & 338.24 & 4.17  & 413.11 & 4.14  & 416.60 & 3.98 & 433.05  \\
    28  &28  &512 &512 &3.45 &9.93  & 346.86 & 7.73  & 445.65 & 7.66  & 449.54 & 7.46 & 461.71  \\
    28  &28  &512 &512 &3.45 &9.92  & 347.48 & 7.73  & 445.53 & 7.67  & 449.31 & 7.48 & 460.42  \\
    14  &14  &512 &512 &0.86 &2.98  & 289.23 & 2.98  & 289.33 & 2.98  & 289.13 & 2.97 & 290.30  \\
    14  &14  &512 &512 &0.86 &2.97  & 289.72 & 2.96  & 290.89 & 2.96  & 290.79 & 2.93 & 293.77  \\
    14  &14  &512 &512 &0.86 &2.97  & 290.40 & 2.97  & 289.72 & 2.96  & 290.60 & 2.96 & 290.99  \\
	\midrule
	\multicolumn{5}{l}{Theoretical peak performance (add/mul) (GOPS)}
        & & 512 & & 648 & & 800 & & 1250 \\
	\multicolumn{5}{l}{Overall performance (GFLOPS)} & & 293.94 & & 353.6 & &  335
                                                   & & 351.86 \\
  \multicolumn{5}{l}{Max fraction of peak sustained} & & 84.1\% & &
    75.5\% & & 72.7\% & & 65.9\% \\

    \bottomrule
  \end{tabularx}
  \caption{Performance breakdown of each layer of VGG-16\label{tab:perf-int8}.
    8 bit multiply and 32 bit accumulate. Batch size is 1. PEs are 4-way vectorized (vPE).
  }
\end{table*}

\subsection{Results and Analysis}
Tables~\ref{tab:util-fp32} and \ref{tab:util-int8} shows the available resources
on the FPGA and the utilization of our design for configurations corresponding
to different numbers of PEs. Note that each PE of the int8 is sort of 4-way
vectorized and we thus use ``vPE'' for it.
LUT usage is very high in fp32 designs as shown in table~\ref{tab:util-fp32}.
Table \ref{tab:pe_usage_fp32} provides breakdown of the LUT usage for fp32 PE,
showing how exactly one of the key resources is being used for different
components of the design --- for a single PE
Table~\ref{tab:perf} and Table~\ref{tab:perf-int8} show performance sustained by
the accelerator for fp32 and int8 (with 32-bit accumulation) respectively across
configurations where we increased the number of PEs.

We now analyze the reasons for the difference in the sustained performance and
the theoretical peak shown in Tables~\ref{tab:perf} and \ref{tab:perf-int8}.
Note that there are broadly two reasons for an under-utilization: (1) an
insufficient amount of memory bandwidth to sustain the computation, and (2) PEs
remaining idle in spite of sufficient memory bandwidth due to a tile
underfitting the dimensions of the processor array (in turn due to problem
sizes). Even in the cases the reason is (1), one could still argue as to whether
a design is exploiting reuse well, i.e., whether there is another design point
that utilizes PEs better while using the {\it same} memory bandwidth. We will
consider this as well.

\paragraph{Performance trend with layer/channel sizes.} The reuse factor on the
output is $C_i * K_x * K_y$. Since $K_x$, $K_y$ are fixed, as we go up the rows
of the tables, we notice that there will not be enough output reuse to be able
to provide the necessary output bandwidth to write out values at the rate at
which they are being produced. This explains the low utilization for small $C_i$
values.  As we go down the rows of the table, we notice the GOPS/GFLOPS
increasing, but they again decrease when $H$, $W$ decrease. Recall the column on
arithmetic intensities in Table~\ref{tab:vgg16}, which indicates that the layers
in the middle have balanced reuse for all three tensors in play. When $H$, $W$
decrease, the degree of weight reuse decreases, and thus the input bandwidth is
not sufficient to provide weights at the required rate (for eg., for the 625 PEs
case with $H = W = 14$, $0.250 * 625 / (14 * 14) * 4 =$ 3 GB/s would be needed).
This is because each core consumes one 32-bit input and weight every cycle and
writes back one 32-bit output.  Hence, one core can have a maximum bandwidth of
1GB/s at 250MHz for input, weight and output each.  We evaluated performance
scaling with only one core but a configuration with fewer PEs per core (at most
14 x 14) and more cores can scale the performance at the expense of extra
bandwidth.

\paragraph{Increase in number of PEs} Now, as we go across the columns of
the tables from left to right, the number of PEs increases, and thus the output
bandwidth requirement also increases even in the presence of optimal output
reuse. One 32-bit value would have to be output every $K_x*K_y*C_i$
cycles for the fp32 design, while it would every $K_x*K_y*C_i / 4$
cycles for the int8 design.  Hence, as we increase the number of PEs, we stop
seeing an improvement in sustained GOPS performance beyond a point. Also, note
that for the same number of PEs, the int8 design has a higher peak GOPS rate
(since each of its PE is a 4-way parallel reduction) and takes one fourth the
number of cycles to generate a 32-bit output. Hence, the output bandwidth
requirement for the int8 design would be higher than fp32 for the same number of
PEs. Like the previous situation, performance can be improved at the expense of
more bandwidth by adding more cores (and reducing PEs per core to half). One
possible strategy can be to tile the output channel dimension and run these
tiles in parallel on the cores. This will double the output bandwidth since now
each core is sending outputs at $1$ GB/s.

\paragraph{Under-utilization due to tiling} As mentioned earlier, tiling could
contribute to an under-utilization of the available compute resources.
``Partial'' tiles do not fully utilize the PE array. For example, consider the
VGG layers with $56 \times 56$ output size.  With 625 PEs, the number of tiles
to perform the convolution will at least be $\lceil{\frac{56*56}{625}} \rceil$,
which is six tiles.  The overall utilization is given by $56*56/(6*625)$ which
is 83.6\%.  This under-utilization happens because the last tile is a partial
one and does not have enough compute to keep all 625 PEs busy.
Table~\ref{tab:perf-int8} shows that the performance of a $56 \times 56\times
256\times 256$ layer is 823.8 GOPs which is at 65.9\% of the machine peak.  Out
of this under-utilization of 34.1\%, 16.4\% is attributed to performance loss
due to tiling. This performance loss can be minimized by using larger batch
sizes which will increase available computation in any tile including partial
tiles.

Overall, we obtain a machine peak of 1.25 TeraOPS with the int8 processor array
with 625 vector vPEs running at 250MHz and with about 70\% resource utilization.
The sustained performance is a good fraction of the machine peak unless it is
limited by inherent reuse due to problem sizes and memory bandwidth that one
core can exploit.

\section{Related Work}
\label{sec:related-work}
There has been an incredible amount of work on building accelerators for CNNs,
and machine learning in general, in recent years. Although our design has
presented
and evaluated as a reconfigurable / FPGA-based one, works that targeted ASICs
are also related to ours, and we thus qualitatively compare with some of these
designs.


A number of deep learning accelerators
have
focused on accelerating matrix-matrix multiplication of certain size matrices. A
CNN or a larger size matrix-matrix multiplication is then built out of mapping
to and composing such smaller matrix-matrix multiplications (matmul). The Google
TPU~\cite{tpu} and the NVDIA GPU's tensor cores~\cite{nvidia-tensor-core} are
prominent ones among such designs, and there are
others~\cite{chen14asplos,liu15asplos} based on BLAS primitives.  Using matmul
as a primitive simplifies the design space, but when used for a convolution
leads to replication of data at some distance from the compute on the chip
(although still on chip). However, a design that does not use matmul as a
primitive brings that replication closer to actual operators (add/multiply) on
the chip. In contrast to approaches that accelerate matmul, our accelerator does
not require an algorithm to be cast into matrix multiplications. It directly
models a convolution, and as such, the reuse of data along the convolution
window happens much closer to the processing elements as opposed to in a on chip
scratchpad. The Google TPU~\cite{tpu} implements a systolic array style
architecture to accelerate matrix matrix multiplications.  The convolution
operation could be mapped to smaller matrix multiplications by tiling and
flattening the input and weight tensors. A local unified buffer caches
intermediate activations for use in the next layer computation. Similarly, there
are other accelerators~\cite{chen14asplos,liu15asplos} that are based on
specialized units to accelerate matrix-vector multiply, and these are are all
more meaningful on an ASIC that is to some extent more programmable via
instructions as opposed being closer to a purer dataflow style design like ours.
As mentioned earlier, this is a trade-off made at the expense of exploiting
reuse at some distance from the actual multipliers and adders, albeit still on
the chip.

Eyeriss~\cite{eyeriss} is a flexible CNN accelerator that uses a two
dimensional array design. It use a dataflow technique that the authors refer to
as row stationary to maximize input, weight and output reuse.  Each processing
element in the two-dimensional array is responsible for a one-dimensional
convolution of an input row and a kernel row to create a row of partial sum
outputs.  The partial sums of multiple PEs are then accumulated to calculate the
output. The PEs are connected to their neighbors in such a way that the inputs,
weights and partial sums are all reused within the PE array.  Eyeriss uses a two
level bus hierarchy to transfer inputs to a set of PEs. The architecture is
easily adapted to different layer shapes like ours. Eyeriss v2~\cite{eyeriss2}
is able to deal with sparsity as well, which we do not address here. In
comparison with Eyeriss, we believe that our interconnection is much simpler,
dealing with a subset of dataflow that Eyeriss v2 deals with. A more direct
comparison to quantitatively evaluate the efficiency of the interconnects or the
final performance is infeasible due to the very different hardware substrates
and processes at play (FPGA vs ASIC).

DnnWeaver~\cite{sharma16micro} provides a template architecture from which a
specialized accelerator is generated. It exploits data reuse via forwarding of
inputs among PEs and dedicated buffers for inputs, weights and outputs. Its
design is composed of multiple processing units, each containing multiple
processsing engines. One key difference between our designs are that DnnWeaver
caches weights locally but PEs in our design do not cache weights.

Multiple FPGA based accelerators have been proposed in the literature.  Zhang et
al.~\cite{zhang15fpga} achieves 61.2 GFLOPs on Alexnet~\cite{alex12nips} with
Virtex7 VX485T FPGA while using 2240 DSP slices.  We achieve higher performance
with 652 DSP slices as shown in Table~\ref{tab:perf} because of higher frequency
of operation. Caffeine~\cite{zhang16iccad} uses a 16 bit fixed point
implementation and achieves 488 GOPs overall. In contrast, we only achieve 351
GOPs using int8 multiply with 32 bit accumulate. This is primarily because
caffeine uses a batch size of 32 whereas we only use a batch size of one.
Increasing the batch size increases weight reuse and would significantly improve
overall performance. TGPA~\cite{wei18iccad} attempts to solve the
underutilization problem due to tensor shape diversity by adopting a
heterogenous architecture. It achieves 1510GOPs of 16 bit fixed point
performance for VGG on a VU9P FPGA while using 4096 DSP blocks. 
In addition to these, multiple other systolic array based designs
\cite{wei17dac,zhang19iscas} have been proposed in the past.  When compared to
these designs, our architecture is different in how it exploits reuse and in the
design of the interconnect.


In addition to the above published works, Xilinx provides DPU (Deep Learning
Processing Unit) accelerator IP~\cite{xilinx-dpu} for acceleration of DNNs on
Zynq-7000 SoC and UltraScale+ MPSoC family of FPGAs. Although a direct
comparison with our work is also not possible, \cite{xilinx-dpu} provides
comprehensive data on end-to-end performance on 8-bit integer quantized
precision  that could be compared in a future work when we are able to report
aggregate end-to-end performance on CNN models in frames per second.  The latter
would require paying attention to a number of other integration issues --- our
focus here has been to evaluate the performance of just the convolution
kernels/layers in greater depth.

Previous work
  \cite{precision2,precision1,precision3} has shown that CNN inference could be
  achieved with low precision arithmetic.  Stripes \cite{stripes} implements a
  bit serial computing and provides a mechanism to make an on-the-fly tradeoff
  between accuracy, performance and power.  In contrast,
  bitfusion~\cite{bitfusion} can dynamically fuse multiple bit-level compute
  elements to match the required precision for computation of each layer.
  Minerva~\cite{minerva} proposes an automated design flow to optimize hardware
  accelerators. It uses data type quantization and operation pruning to reduce
  power consumption. EIE~\cite{eie}, SCNN~\cite{parashar17arxiv} and
  Cnvlutin~\cite{cnvlutin} exploit sparsity in weights and input feature map to
  skip computations and improve overall performance.  Techniques such as deep
  deep compression~\cite{deep_compression} complement these accelerators by
  increasing the sparsity of the weight matrix and reducing the required
  precision without impacting overall accuracy of the model. All of these
  techniques are orthogonal to our approach.

\section{Conclusions}
\label{sec:conclusions} We proposed an FPGA-based accelerator design to execute
convolutional neural networks while exploiting reuse along all dimensions. Our
accelerator core, which is a 1-d systolic array of processing elements, is
highly flexible and avoids reconfiguration while allowing high utilization for
arbitrary aspect ratio tiles of the larger layer dimensions. The design achieved
a high clock frequency even with close to maximal utilization.  We described how
the accelerator could be leveraged transparently in a deep learning programming
model like TensorFlow with the necessary software codesign. Experimental
evaluation on a real system with a PCI-express based FPGA accelerator
demonstrated the effectiveness of the accelerator in sustaining as high a
fraction of the peak as reuse and memory bandwidth would have allowed. We intend
to make our entire design open and publicly available.

\section*{Acknowledgements}
We would like to gratefully acknowledge the Science and Engineering Research
Board (SERB), Government of India, for funding this research work in part
through a grant under its EMR program (EMR/2016/008015).

\balance
\bibliographystyle{IEEEtranS}
\bibliography{bibfile.bib}

\begin{thebibliography}{10}
\providecommand{\url}[1]{#1}
\csname url@samestyle\endcsname
\providecommand{\newblock}{\relax}
\providecommand{\bibinfo}[2]{#2}
\providecommand{\BIBentrySTDinterwordspacing}{\spaceskip=0pt\relax}
\providecommand{\BIBentryALTinterwordstretchfactor}{4}
\providecommand{\BIBentryALTinterwordspacing}{\spaceskip=\fontdimen2\font plus
\BIBentryALTinterwordstretchfactor\fontdimen3\font minus
  \fontdimen4\font\relax}
\providecommand{\BIBforeignlanguage}[2]{{%
\expandafter\ifx\csname l@#1\endcsname\relax
\typeout{** WARNING: IEEEtranS.bst: No hyphenation pattern has been}%
\typeout{** loaded for the language `#1'. Using the pattern for}%
\typeout{** the default language instead.}%
\else
\language=\csname l@#1\endcsname
\fi
#2}}
\providecommand{\BIBdecl}{\relax}
\BIBdecl

\bibitem{cnvlutin}
\BIBentryALTinterwordspacing
J.~Albericio, P.~Judd, T.~Hetherington, T.~Aamodt, N.~E. Jerger, and
  A.~Moshovos, ``Cnvlutin: Ineffectual-neuron-free deep neural network
  computing,'' in \emph{Proceedings of the 43rd International Symposium on
  Computer Architecture}, ser. ISCA '16.\hskip 1em plus 0.5em minus 0.4em\relax
  Piscataway, NJ, USA: IEEE Press, 2016, pp. 1--13. [Online]. Available:
  \url{https://doi.org/10.1109/ISCA.2016.11}
\BIBentrySTDinterwordspacing

\bibitem{precision2}
S.~Anwar, K.~Hwang, and W.~Sung, ``Fixed point optimization of deep
  convolutional neural networks for object recognition,'' in \emph{2015 IEEE
  International Conference on Acoustics, Speech and Signal Processing
  (ICASSP)}, April 2015, pp. 1131--1135.

\bibitem{chen14asplos}
T.~Chen, Z.~Du, N.~Sun, J.~Wang, C.~Wu, Y.~Chen, and O.~Temam, ``Diannao: A
  small-footprint high-throughput accelerator for ubiquitous
  machine-learning,'' in \emph{Proceedings of the 19th International Conference
  on Architectural Support for Programming Languages and Operating Systems},
  ser. ASPLOS '14.\hskip 1em plus 0.5em minus 0.4em\relax New York, NY, USA:
  ACM, 2014, pp. 269--284.

\bibitem{eyeriss}
Y.~{Chen}, J.~{Emer}, and V.~{Sze}, ``Eyeriss: A spatial architecture for
  energy-efficient dataflow for convolutional neural networks,'' in \emph{2016
  ACM/IEEE 43rd Annual International Symposium on Computer Architecture
  (ISCA)}, June 2016, pp. 367--379.

\bibitem{r-cnn}
\BIBentryALTinterwordspacing
R.~B. Girshick, J.~Donahue, T.~Darrell, and J.~Malik, ``Rich feature
  hierarchies for accurate object detection and semantic segmentation,''
  \emph{CoRR}, vol. abs/1311.2524, 2013. [Online]. Available:
  \url{http://arxiv.org/abs/1311.2524}
\BIBentrySTDinterwordspacing

\bibitem{goto2008toms}
\BIBentryALTinterwordspacing
K.~Goto and R.~A. v.~d. Geijn, ``Anatomy of high-performance matrix
  multiplication,'' \emph{ACM Trans. Math. Softw.}, vol.~34, no.~3, pp.
  12:1--12:25, May 2008. [Online]. Available:
  \url{http://doi.acm.org/10.1145/1356052.1356053}
\BIBentrySTDinterwordspacing

\bibitem{precision1}
\BIBentryALTinterwordspacing
S.~Gupta, A.~Agrawal, K.~Gopalakrishnan, and P.~Narayanan, ``Deep learning with
  limited numerical precision,'' in \emph{Proceedings of the 32Nd International
  Conference on International Conference on Machine Learning - Volume 37}, ser.
  ICML'15.\hskip 1em plus 0.5em minus 0.4em\relax JMLR.org, 2015, pp.
  1737--1746. [Online]. Available:
  \url{http://dl.acm.org/citation.cfm?id=3045118.3045303}
\BIBentrySTDinterwordspacing

\bibitem{eie}
\BIBentryALTinterwordspacing
S.~Han, X.~Liu, H.~Mao, J.~Pu, A.~Pedram, M.~A. Horowitz, and W.~J. Dally,
  ``Eie: Efficient inference engine on compressed deep neural network,''
  \emph{SIGARCH Comput. Archit. News}, vol.~44, no.~3, pp. 243--254, Jun. 2016.
  [Online]. Available: \url{http://doi.acm.org/10.1145/3007787.3001163}
\BIBentrySTDinterwordspacing

\bibitem{deep_compression}
S.~Han, H.~Mao, and W.~J. Dally, ``Deep compression: Compressing deep neural
  networks with pruning, trained quantization and huffman coding,''
  \emph{International Conference on Learning Representations (ICLR)}, 2016.

\bibitem{resnet}
\BIBentryALTinterwordspacing
K.~He, X.~Zhang, S.~Ren, and J.~Sun, ``Deep residual learning for image
  recognition,'' \emph{CoRR}, vol. abs/1512.03385, 2015. [Online]. Available:
  \url{http://arxiv.org/abs/1512.03385}
\BIBentrySTDinterwordspacing

\bibitem{xilinx-dpu}
X.~Inc, ``{DPU} for convolutional neural network v3.0,'' 2019.

\bibitem{riffa}
M.~{Jacobsen} and R.~{Kastner}, ``Riffa 2.0: A reusable integration framework
  for fpga accelerators,'' in \emph{2013 23rd International Conference on Field
  programmable Logic and Applications}, Sep. 2013, pp. 1--8.

\bibitem{tpu}
\BIBentryALTinterwordspacing
N.~P. Jouppi, C.~Young, N.~Patil, D.~Patterson, G.~Agrawal, R.~Bajwa, S.~Bates,
  S.~Bhatia, N.~Boden, A.~Borchers, R.~Boyle, P.-l. Cantin, C.~Chao, C.~Clark,
  J.~Coriell, M.~Daley, M.~Dau, J.~Dean, B.~Gelb, T.~V. Ghaemmaghami,
  R.~Gottipati, W.~Gulland, R.~Hagmann, C.~R. Ho, D.~Hogberg, J.~Hu, R.~Hundt,
  D.~Hurt, J.~Ibarz, A.~Jaffey, A.~Jaworski, A.~Kaplan, H.~Khaitan,
  D.~Killebrew, A.~Koch, N.~Kumar, S.~Lacy, J.~Laudon, J.~Law, D.~Le, C.~Leary,
  Z.~Liu, K.~Lucke, A.~Lundin, G.~MacKean, A.~Maggiore, M.~Mahony, K.~Miller,
  R.~Nagarajan, R.~Narayanaswami, R.~Ni, K.~Nix, T.~Norrie, M.~Omernick,
  N.~Penukonda, A.~Phelps, J.~Ross, M.~Ross, A.~Salek, E.~Samadiani, C.~Severn,
  G.~Sizikov, M.~Snelham, J.~Souter, D.~Steinberg, A.~Swing, M.~Tan,
  G.~Thorson, B.~Tian, H.~Toma, E.~Tuttle, V.~Vasudevan, R.~Walter, W.~Wang,
  E.~Wilcox, and D.~H. Yoon, ``In-datacenter performance analysis of a tensor
  processing unit,'' in \emph{Proceedings of the 44th Annual International
  Symposium on Computer Architecture}, ser. ISCA '17.\hskip 1em plus 0.5em
  minus 0.4em\relax New York, NY, USA: ACM, 2017, pp. 1--12. [Online].
  Available: \url{http://doi.acm.org/10.1145/3079856.3080246}
\BIBentrySTDinterwordspacing

\bibitem{stripes}
P.~Judd, J.~Albericio, T.~Hetherington, T.~M. Aamodt, and A.~Moshovos,
  ``Stripes: Bit-serial deep neural network computing,'' in \emph{2016 49th
  Annual IEEE/ACM International Symposium on Microarchitecture (MICRO)}, Oct
  2016, pp. 1--12.

\bibitem{maeri}
\BIBentryALTinterwordspacing
H.~Kwon, A.~Samajdar, and T.~Krishna, ``Maeri: Enabling flexible dataflow
  mapping over dnn accelerators via reconfigurable interconnects,'' in
  \emph{Proceedings of the Twenty-Third International Conference on
  Architectural Support for Programming Languages and Operating Systems}, ser.
  ASPLOS '18.\hskip 1em plus 0.5em minus 0.4em\relax New York, NY, USA: ACM,
  2018, pp. 461--475. [Online]. Available:
  \url{http://doi.acm.org/10.1145/3173162.3173176}
\BIBentrySTDinterwordspacing

\bibitem{lavin2015fast}
A.~Lavin and S.~Gray, ``Fast algorithms for convolutional neural networks,''
  2015.

\bibitem{precision3}
\BIBentryALTinterwordspacing
D.~D. Lin, S.~S. Talathi, and V.~S. Annapureddy, ``Fixed point quantization of
  deep convolutional networks,'' in \emph{Proceedings of the 33rd International
  Conference on International Conference on Machine Learning - Volume 48}, ser.
  ICML'16.\hskip 1em plus 0.5em minus 0.4em\relax JMLR.org, 2016, pp.
  2849--2858. [Online]. Available:
  \url{http://dl.acm.org/citation.cfm?id=3045390.3045690}
\BIBentrySTDinterwordspacing

\bibitem{liu15asplos}
D.~Liu, T.~Chen, S.~Liu, J.~Zhou, S.~Zhou, O.~Teman, X.~Feng, X.~Zhou, and
  Y.~Chen, ``Pudiannao: A polyvalent machine learning accelerator,'' in
  \emph{Proceedings of the Twentieth International Conference on Architectural
  Support for Programming Languages and Operating Systems}, ser. ASPLOS
  '15.\hskip 1em plus 0.5em minus 0.4em\relax ACM, 2015, pp. 369--381.

\bibitem{mlir}
``{MLIR}: Multi-level intermediate representation,'' 2019,
  https://github.com/tensorflow/mlir.

\bibitem{nvidia-tensor-core}
``{NVIDIA} {GPU} tensor cores,'' 2019,
  https://developer.nvidia.com/tensor-cores.

\bibitem{parashar17arxiv}
\BIBentryALTinterwordspacing
A.~Parashar, M.~Rhu, A.~Mukkara, A.~Puglielli, R.~Venkatesan, B.~Khailany,
  J.~S. Emer, S.~W. Keckler, and W.~J. Dally, ``{SCNN:} an accelerator for
  compressed-sparse convolutional neural networks,'' \emph{CoRR}, vol.
  abs/1708.04485, 2017. [Online]. Available:
  \url{http://arxiv.org/abs/1708.04485}
\BIBentrySTDinterwordspacing

\bibitem{alex12nips}
F.~Pereira, C.~J.~C. Burges, L.~Bottou, and K.~Q. Weinberger, Eds.,
  \emph{ImageNet Classification with Deep Convolutional Neural Networks}.\hskip
  1em plus 0.5em minus 0.4em\relax Curran Associates, Inc., 2012.

\bibitem{minerva}
B.~Reagen, P.~Whatmough, R.~Adolf, S.~Rama, H.~Lee, S.~K. Lee, J.~M.
  Hernández-Lobato, G.~Wei, and D.~Brooks, ``Minerva: Enabling low-power,
  highly-accurate deep neural network accelerators,'' in \emph{2016 ACM/IEEE
  43rd Annual International Symposium on Computer Architecture (ISCA)}, June
  2016, pp. 267--278.

\bibitem{yolo}
\BIBentryALTinterwordspacing
J.~Redmon, S.~K. Divvala, R.~B. Girshick, and A.~Farhadi, ``You only look once:
  Unified, real-time object detection,'' \emph{CoRR}, vol. abs/1506.02640,
  2015. [Online]. Available: \url{http://arxiv.org/abs/1506.02640}
\BIBentrySTDinterwordspacing

\bibitem{faster-r-cnn}
\BIBentryALTinterwordspacing
S.~Ren, K.~He, R.~B. Girshick, and J.~Sun, ``Faster {R-CNN:} towards real-time
  object detection with region proposal networks,'' \emph{CoRR}, vol.
  abs/1506.01497, 2015. [Online]. Available:
  \url{http://arxiv.org/abs/1506.01497}
\BIBentrySTDinterwordspacing

\bibitem{sayres19opthalmology}
R.~Sayres, A.~Taly, E.~Rahimy, K.~Blumer, D.~Coz, N.~Hammel, J.~Krause,
  A.~Narayanaswamy, Z.~Rastegar, D.~Wu, S.~Xu, S.~Barb, A.~Joseph, M.~Shumski,
  J.~Smith, A.~B. Sood, G.~S. Corrado, L.~Peng, and D.~R. Webster, ``Using a
  deep learning algorithm and integrated gradients explanation to assist
  grading for diabetic retinopathy,'' \emph{Ophthalmology}, vol. 126, no.~4,
  pp. 552 -- 564, 2019.

\bibitem{sharma16micro}
H.~{Sharma}, J.~{Park}, D.~{Mahajan}, E.~{Amaro}, J.~K. {Kim}, C.~{Shao},
  A.~{Mishra}, and H.~{Esmaeilzadeh}, ``From high-level deep neural models to
  fpgas,'' pp. 1--12, Oct 2016.

\bibitem{bitfusion}
H.~Sharma, J.~Park, N.~Suda, L.~Lai, B.~Chau, V.~Chandra, and H.~Esmaeilzadeh,
  ``Bit fusion: Bit-level dynamically composable architecture for accelerating
  deep neural network,'' in \emph{{ISCA}}.\hskip 1em plus 0.5em minus
  0.4em\relax {IEEE} Computer Society, 2018, pp. 764--775.

\bibitem{simonyan2014arxiv}
K.~Simonyan and A.~Zisserman, ``Very deep convolutional networks for
  large-scale image recognition,'' 2014.

\bibitem{vanzee2015toms}
\BIBentryALTinterwordspacing
F.~G. Van~Zee and R.~A. van~de Geijn, ``Blis: A framework for rapidly
  instantiating blas functionality,'' \emph{ACM Trans. Math. Softw.}, vol.~41,
  no.~3, pp. 14:1--14:33, Jun. 2015. [Online]. Available:
  \url{http://doi.acm.org/10.1145/2764454}
\BIBentrySTDinterwordspacing

\bibitem{wei18iccad}
X.~{Wei}, Y.~{Liang}, X.~{Li}, C.~H. {Yu}, P.~{Zhang}, and J.~{Cong}, ``Tgpa:
  Tile-grained pipeline architecture for low latency cnn inference,'' in
  \emph{2018 IEEE/ACM International Conference on Computer-Aided Design
  (ICCAD)}, Nov 2018, pp. 1--8.

\bibitem{resnext}
\BIBentryALTinterwordspacing
S.~Xie, R.~B. Girshick, P.~Doll{\'{a}}r, Z.~Tu, and K.~He, ``Aggregated
  residual transformations for deep neural networks,'' \emph{CoRR}, vol.
  abs/1611.05431, 2016. [Online]. Available:
  \url{http://arxiv.org/abs/1611.05431}
\BIBentrySTDinterwordspacing

\bibitem{wei17dac}
{Xuechao Wei}, {Cody Hao Yu}, {Peng Zhang}, {Youxiang Chen}, {Yuxin Wang}, {Han
  Hu}, {Yun Liang}, and J.~{Cong}, ``Automated systolic array architecture
  synthesis for high throughput cnn inference on fpgas,'' in
  \emph{ACM/EDAC/IEEE Design Automation Conference (DAC)}, June 2017, pp. 1--6.

\bibitem{eyeriss2}
{Y.-H. Chen, J. Emer, V. Sze}, ``{Eyeriss v2: A Flexible and High-Performance
  Accelerator for Emerging Deep Neural Networks},'' in \emph{{arXiv}}, {2018}.

\bibitem{zhang16iccad}
C.~{Zhang}, {Zhenman Fang}, {Peipei Zhou}, {Peichen Pan}, and {Jason Cong},
  ``Caffeine: Towards uniformed representation and acceleration for deep
  convolutional neural networks,'' in \emph{IEEE/ACM International Conference
  on Computer-Aided Design (ICCAD)}, Nov 2016, pp. 1--8.

\bibitem{zhang15fpga}
C.~Zhang, P.~Li, G.~Sun, Y.~Guan, B.~Xiao, and J.~Cong, ``Optimizing
  {FPGA}-based accelerator design for deep convolutional neural networks,'' in
  \emph{Proceedings of the 2015 ACM/SIGDA International Symposium on
  Field-Programmable Gate Arrays}, ser. FPGA '15, 2015, pp. 161--170.

\bibitem{zhang19iscas}
J.~{Zhang}, W.~{Zhang}, G.~{Luo}, X.~{Wei}, Y.~{Liang}, and J.~{Cong},
  ``Frequency improvement of systolic array-based cnns on {FPGA}s,'' in
  \emph{2019 IEEE International Symposium on Circuits and Systems (ISCAS)}, May
  2019, pp. 1--4.

\end{thebibliography}

\end{document}